# ANCIENT ORIGINS OF A MODERN ANTHROPIC COSMOLOGICAL ARGUMENT

**Dr. Milan M. Ćirković**

*Astronomical Observatory Belgrade*

*Volgina 7*

*11000 Belgrade*

*YUGOSLAVIA*

*e-mail:* `arioch@eunet.yu`

# ANCIENT ORIGINS OF A MODERN ANTHROPIC COSMOLOGICAL ARGUMENT

**Abstract.** Ancient origins of a modern anthropic argument against cosmologies involving infinite series of past events are considered. It is shown that this argument—which in modern times has been put forward by distinguished cosmologists like Paul C. W. Davies and Frank J. Tipler—originates in pre-Socratic times and is implicitly present in the cyclical cosmology of Empedocles. There are traces of the same line of reasoning throughout the ancient history of ideas, and the case of a provocative statement of Thucydides is briefly analyzed. Moreover, the anthropic argument has been fully formulated in the epic of Lucretius, confirming it as the summit of ancient cosmology. This is not only of historical significance but presents an important topic for the philosophy of cosmology provided some of the contemporary inflationary models, particularly Linde's chaotic inflation, turn out to be correct.

## 1. Introduction: Davies-Tipler argument

The simplest division of all cosmologies is into two broad classes: those postulating the eternal universe and those which postulate some origin of the universe, or at least the part of it that cosmologists are currently inhabiting. Eternal universes (and here by eternal I mean either those with no temporal beginning or end or those with no beginning only) are the only ones which could pretend to adopt some sort of stationarity, a condition which is of singular importance in many branches of physics (among other issues because the law of energy conservation is closely connected with a translational symmetry of time), and which is certainly seen as greatly simplifying the solution of specific problems everywhere. For a long period of time, after the religious dogma about Creation in 4004 BC (or any other specific date) was abandoned, the universe has been considered eternal, although great minds, such as Newton's, began to perceive some of the difficulties associated with such a proposition (e.g. North 1965). The resistance to any opposing view (which eventually became what is today dubbed the standard cosmology) was exceedingly strong during most of the nineteenth and the early twentieth century. It is epitomized in the words of one of the pioneers of modern astrophysics, Sir Arthur Eddington, who in his authoritative monograph *The Nature of the Physical World* wrote: "As a scientist, I simply do not believe that the universe began with a bang."[1] From the end of the Middle Ages until Hubble's observational revolution in the third decade of the twentieth century, the stationary worldview has been in one way or another the dominant one. This explains, among other issues, the dramatic reaction of most of the scientific community, including Lord Kelvin, Holmes, Eddington, Crookes, Jeans and others, to the discoveries of Clausius, Boltzmann and other thermodynamicists, implying a unidirectional flow of time and physical change. Interestingly enough, even during this epoch the idea—today one of most investigated issues in physics—that the thermodynamical arrow of time originates in cosmology, has occasionally surfaced (Steckline 1983; Price 1996, and references therein).

The power of a stationary alternative to the evolutionary models of the universe has been reiterated in particularly colorful form during the great cosmological controversy in late 1940-ies, 1950-ies and early 1960-ies (Kragh 1996). Although during this period of conflict between the Big Bang and the classical steady state theories numerous and very heterogeneous arguments appeared on both sides of the controversy, the argument based on the anthropic selection effect was only explicitly formulated a decade after the disagreements ended. As it is well known, the debate ceased when empirical arguments persuaded by far the largest part of the cosmological community that a universe of finite age is the only empirically acceptable concept.[2] However, the argument based on the anthropic principle has been further developed during the 1980s and has gained relevance in a new and developing field of quantum cosmology (together with other aspects of anthropic reasoning). This brief note is dedicated to investigation of its origin in the ancient philosophy of nature, while the detailed consideration of its range, scope and various versions is forthcoming (manuscript in preparation).

The modern version of the anthropic argument against the past infinite series of events (or the past temporal infinity in relationist terms; see the discussion below) has appeared in a short notice by Paul C. W. Davies appearing in *Nature* in June 1978 (Davies 1978). In this succinct critique of the Ellis et al. (1978) static cosmological model Davies points out that

> there is also the curious problem of why, if the Universe is infinitely old and life is concentrated in our particular corner of the cosmos, it is not inhabited by technological communities of unlimited age.

The same idea has been further developed and put on a mathematical footing by Tipler (1982). As claimed by Barrow and Tipler (1986) in their encyclopaedic monograph on anthropic principles, this is historically the first instance in which an anthropic argument has been used against cosmology containing the past temporal infinity. As we shall see in the rest of this study, this claim is only partially correct, since the thinkers in antiquity have been aware of a similar argumentation. However, it is indeed fascinating that the same argument had not been considered earlier in the course of XX century. The suprise is strengthened by the fact that cosmologies postulating an infinite past in scientific or half-scientific form have existed since the very dawn of science. In addition, since ancient times a belief in the existence of other **inhabited** worlds has also been present, in one form or another.[3] Today, the scepticism sometimes encountered against this mode of thinking is even stranger, when various (and, at least in some cases, not quite inexpensive) SETI projects testify to the reasonable degree of belief in the existence of technological civilizations other than the human one. Their technological nature (the same one which produces the problem Davies wrote about) is a *conditio sine qua non* of any sensible SETI enterprise. In this short note, we shall try to recall some of the instances this argument has surfaced in the ancient cosmological thought, while leaving the deep tracing of its elements and possibly a wide survey to a subsequent work.

## 2. Empedocles' uniformitarianism and reductionism

An ancient echo of this type of argumentation can be recognized in the surviving fragments of some of the most distinguished Hellenic philosophers of nature. From our point of view especially interesting is the cyclic cosmology of Empedocles of Acragas (VI-V century BC), in which the universe is eternal,[4] consisting of the internally immutable four classic elements, as well as two opposing forces (Love and Strife, i.e. attractive and repulsive interactions). The cyclic motion of matter in the universe is governed by the change in relative intensities of two interactions (see the excellent discussion in O'Brien 1969). It is interesting to note that Empedocles' cosmology is **uniformitarian**, in the sense that all six basic constituents (four elements and two forces between them) are present in each instant of time in accordance with the eternal principles of mutual exchange. In some of the surviving fragments, Empedocles implies that although this uniformitarianism may seem counterintuitive, as we see things coming into being and vanishing, this is just our special perspective (today we would say **anthropocentrism**) and not the inherent state of nature.[5] This is strikingly similar to the uniformitarian notions present in some of the most authoritative cosmological models of the twentieth century, notably the classical steady state theory (Balashov 1994). The connection is strikingly relevant when the fact that the classical steady state theory entails an infinite past is taken into account.

However, the most interesting aspect of this cosmological picture is what occurs within each individual great cosmic cycle. Probably the most lasting and controversial legacy of Empedoclean cosmology is his assertion that biological and even anthropological evolution are **inherent, necessary and inseparable** parts of the global cosmological evolution (Guthrie 1969). Thus, speaking on the four elements, he states

> For out of these have sprung all things that were and are and shall be – trees and men and women, beasts and birds and the fishes that dwell in the waters, yea, and the gods that live long lives and are exalted in honor.
> For these things are what they are; but, running through one another, they take different shapes – so much does mixture change them.[6]

However, if we accept this view—which we shall call an "Empedoclean picture" in the further text—that biological evolution and the appearance of consciousness and intelligence are contingent upon cosmological processes, the eternal universe of Empedocles faces the same kind of problem as that of modern stationary cosmologies like the classical steady state theory or the one of Ellis et al. criticized by Davies. Why then, in the supposed infinity of time, are "men and women, beasts and birds" of finite, and relatively small, age? Empedocles may have perceived this himself (his mode of thinking, and even his theory of metempsychosis, were closer to the modern anthropic mode of thinking than most of the later physicists and philosophers), and he evades the problem in the only natural way he can: by postulating two singular states in the beginning and in the middle of each of his great cycles. These singular states are moments (in the absolute time!) of complete dominance of either Love (an ancient equivalent of the modern initial and/or final singularities) or Strife (no true equivalent, but similar to the modern version of heat death in the ever-expanding cosmological models; see, for instance, Davies 1994). In these states the life, with its complex organizational structure, is impossible and therefore they serve as *termini* for the duration of any individual history of life and intelligence. The maximal duration of any form of life and/or intelligence is determined exclusively by cosmological laws. Therefore, there are no arbitrarily old beings, and anthropic argument is inapplicable.

It is worth noting that the Empedoclean reductionist picture of the relationship between biological and psychological processes on one hand, and physical and cosmological processes on the other, has become quite common in the ancient philosophical thought after Empedocles. It is also present, for instance, in cosmologies postulating finite age of the universe, or at least a finite duration of world histories, such as in Anaxagoras' system. According to the testimony of Diodorus (I 7, 7), Euripides has, in his lost tragedy *Melanippa*, described—clearly under the influence of his teacher Anaxagoras—the rise of plants, animals and humans as an ultimate consequence of separation of the Heavens and the Earth from their primordial unity; which is another suprisingly modern picture. With the rise of Socrates, and subsequently Platonic and Aristotelian philosophy, and in particular during the age of faith, this line of thinking became discontinued; in a sense it has only inherited worthy successors in the modern thought contained in philosophical considerations of both quantum mechanics and cosmology (e.g. Schrödinger 1944; Barrow and Tipler 1986; Smolin 1997). We can not treat these reissues of the Empedoclean picture in the course of this study. However, it is worth noticing that the problems facing such contingency of biological upon cosmological processes have also been noted in antiquity by several famous authors.

## 3. Repetitions in antiquity

In the very first chapter of the immortal history of Thucydides, there is a famous statement that before his time—i.e. about 450 BC—nothing of importance (συ μεγαλα γενεσθαι) had happened in history. This startling statement has been correctly called "outrageous" by Spengler, and used to demonstrate the essentially mythological character of ancient Greek historiography (Spengler 1918; see also Cornford 1965). It may indeed be outrageous from the modern perspective, but it does motivate a set of deeper questions, ultimately dealing with cosmology. The fact that Thucydides did not know (or did not care to know) previous historical events does not change the essential perception of **finiteness** of human history inseparable from the Greek thought. This property starkly conflicts with the notion of an **eternal continuously existent** world, as it was presented in both modern and ancient cultures. Obviously, it is irrelevant which exact starting point we choose for unfolding historical events. In any case, the number of these events is finite, and the timespan considered small even compared to the specific astronomical timescales (some of which, like the precession period of equinoxes, were known in the classical antiquity, as is clear from the discussion in *Timaeus*), not to mention anything about a past temporal infinity. Although there was no scientific archaeology in the ancient world, it was as natural then as it is now to expect hypothetical previous civilizations inhabiting Oikumene to leave some traces—in fact, an infinite number of traces for an eternally existent Oikumene! There are indications that pre-Socratic thinkers have been aware of the incompatibility of this "Thucydidean" finiteness of historical past with the eternal nature of the world. We have already mentioned the solution (periodic singular states) proposed by Empedocles himself. Even earlier, in the fragmentary accounts of the cosmology of Anaximandros, one may note that he proposed an evolutionary origin of humankind in some finite moment in the past, parallel with his basic postulate of separation of different worlds from *apeiron* and their subsequent returning to it.[7] In Anaxagoras' worldview, there is a famous tension between the eternity of the world's constituents and the finite duration of **movement** (and, therefore, relational time) in the world. In the same time, it seems certain that Anaxagoras, together with Anaximandros and Empedocles, was an early proponent of the evolutionary view, at least regarding the origin of humankind (Guthrie 1969).

Finally, an almost modern formulation of the anthropic argument against the past temporal infinity has been made in Roman times by Lucretius, who in Book V of his famous poem *De Rerum Natura* wrote the following intriguing verses:

> Besides all this,
> If there had been no origin-in-birth
> Of lands and sky, and they had ever been
> The everlasting, why, ere Theban war
> And obsequies of Troy, have other bards
> Not also chanted other high affairs?
> Whither have sunk so oft so many deeds
> Of heroes? Why do those deeds live no more,
> Ingrafted in eternal monuments
> Of glory? Verily, I guess, because
> The Sun is new, and of a recent date
> The nature of our universe, and had
> Not long ago its own exordium.[8]

For highly scientific-minded Lucretius, the shortness of human history **is** very strange on the face of hypothesis of the eternal existence of the world. Although the references to "eternal monuments" and "other bards" may sound naive, it is clear that he had in mind any form of transmission of information from the past to the present; and an infinite amount of information from an infinite past. His empirical assessment of the surrounding world clearly shows the absence of such information. Therefore, an explanation is needed. The simplest explanation, as Lucretius was highly aware, is to treat the argument as *reductio ad absurdum* of the starting hypothesis (eternal nature of the world) and to assume that the world is of finite—and relatively small—age.

The depth of Lucretius' thought in this passage is almost amazing, especially when the historical blindness of subsequent generations to this same argumentation is taken into account. The Lucretius' argument applies to the classical Newtonian universe of infinite age, as well as to modern stationary alternatives to the evolutionary cosmology. It emphasizes the **technological** nature of possible evidence ("ingrafted in… monuments"). This is exactly what modern cosmologists Davies and Tipler have had in mind when constructing the anthropic argument in order to refute the eternal cosmologies of our epoch. Lucretius' monuments play essentially the same role as Tipler's von Neumann probes sent by advanced intelligent communities (Tipler 1982). Thus, Lucretius undoubtedly presents a summit of ancient philosophical discussion of the question of the age of the world.

**4. Lessons**

We have seen another instance surprising modernity of views and debates of the classical world in respect to the issues of (i) the age of the universe, and (ii) the place of intelligent observers in it. However, most of lessons of it seemed to be forgotten in the course of history, and it is therefore not surprising to find many fallacies and misleading statements in the modern sources on these same questions. Notably, the Empedoclean issue whether cosmological evolution leads to intelligence and consciousness seem to be abused in what one may term the cosmological double standard towards the conditions for existence of intelligent observer. On the one hand, scientists do regard our presence (and presence of any other intelligent observers which may exist in the universe) as purely incidental, and hesitate to draw strong conclusions about nature from the facts of our existence (e.g. Pagels 1998). All resistance encountered by the anthropic principles testifies on that. On the other hand, we are often asked to *a priori* assume that life, intelligence and consciousness are of natural origin, and that we need not invoke any supranatural (or even just non-physical) causes in explanation of their appearance in the universe. This double standard dealing with the inference or non-inference from the emergence of intelligent observers is closely connected with the temporal double standard, which is the source of many fallacious statements concerning the temporal asymmetry of the physical world (Price 1996). The connection becomes more visible when we take into account the almost trivial conclusion, explicitly formulated and defended by Dyson in his classical paper, founding the young discipline of physical eschatology (Dyson 1979):

> It is impossible to calculate in detail the long-range future of the universe without including the effects of life and intelligence. It is impossible to calculate the capabilities of life and intelligence without touching, at least peripherally, philosophical questions. If we are to examine how intelligent life may be able to guide the physical development of the universe for its own purposes, we cannot altogether avoid considering what the values and purposes of intelligent life may be. But as soon as we mention the words value and purpose, we run into one of the most firmly entrenched taboos of twentieth-century science.

Future of the universe containing life and intelligence is **essentially** different from the past of the same universe in which there were no such forms of complex organization of matter.[9] The Empedoclean picture of continuity between physical and biological evolution implies a form of temporal asymmetry and contradicts the crude atemporal interpretations based on preserving the mind-matter dualism. This raises a host of issues dealing with the impact of complexification, as manifested through biological and subsequent psychological evolution on the universe as a whole, as well as issues in the philosophy of mind, which we cannot discuss here. Whether a more sophisticated atemporal description is capable of accounting for these anthropic restrictions, remains to be seen.

    We have not traced the origination of the Davies-Tipler argument in depth, nor put it in the wider context of ancient cosmologies. This represents an entirely different, vastly more difficult enterprise. However, it is our modest hope that we have demonstrated freshness, novelty and relevance of ideas of ancient thinkers in the interplay with some of the most

active areas of modern scientific and philosophical research, such as theoretical cosmology and philosophy of time.

The core lesson of the entire case of the anthropic argument against cosmologies containing past temporal infinities is, however, located on a deep epistemological level. As a side effect of both the Copernican revolution and the Cartesian dualism, the implicit rejection of the pre-Socratic picture of the inseparability of the cosmological, biological and anthropological domains led to an inevitable delay in noticing a powerful and specific cosmological argument. Further discussions on this topic, as well as further discussions of the future of **physical** universe, will have to **explicitly** take into account the existence and activities of intelligent observers. This will manifest itself not only in retrodictions about the cosmological past, as the original anthropic argument of Dicke and Carter has been traditionally used, but also through the predictive aspect of cosmology. These physical eschatological considerations will necessarily be of multidisciplinary character, so desirable in this latest epoch of development of our picture of the universe. In this respect, reinvestigation and reevaluation of the ancient sources of modern cosmology will certainly be seen as noble and rewarding endevoar.

**References**

Adams, F. C. and Laughlin, G. 1997, *Rev. Mod. Phys.* **69**, 337.

Balashov, Yu. 1994, *Studies in History and Philosophy of Science* **25B**, 933.

Barrow, J. D. and Tipler, F. J. 1986, *The Anthropic Cosmological Principle* (Oxford
  University Press, New York).

Burnet, J. 1908, *Early Greek Philosophy* (Adam and Charles Black, London).

Cornford, F. 1965, *Thucydides Mythistoricus* (Greenwood Press Publishers, New York).

Davies, P. C. W. 1978, *Nature* **273**, 336.

Davies, P. C. W. 1994, *The Last Three Minutes* (Basic Books, New York).

Diels, H. 1983, *Presocratic Fragments* (Naprijed, Zagreb).

Dyson, F. 1979, *Rev. Mod. Phys.* **51**, 447.

Eddington, A. S. 1928, *The Nature of the Physical World* (Cambridge University
  Press, London).

Ellis, G. F. R., Maartens, R. and Nel, S. D. 1978, *Mon. Not. R. astr. Soc.* **184**, 439.

Fairbanks, A. 1898, *The First Philosophers of Greece* (K. Paul, Trench & Trubner, London).

Guthrie, W. K. C. 1969, *A History of Greek Philosophy II* (Cambridge University Press, London).

Kragh, H. 1996, *Cosmology and Controversy* (Princeton University Press, Princeton).

Lucretius 1997, *On the Nature of Things* (translated by William E. Leonard, e-text version, Project Gutenberg, Urbana).

North, J. 1965, *The Measure of the Universe: A History of Modern Cosmology* (Oxford University Press, London).

O'Brien, D. 1969, *Empedocles' Cosmic Cycle* (Cambridge University Press, Cambridge).

Pagels, H. R. 1998, in *Modern Cosmology and Philosophy*, ed. by J. Leslie (Prometheus Books, New York), p. 180.

Peebles, P. J. E. 1993, *Principles of Physical Cosmology* (Princeton University Press, Princeton).

Price, H. 1996, *Time's Arrow and Archimedes' Point* (Oxford University Press, Oxford).

Schrödinger, E. 1944, *What is Life?* (Cambridge University Press, Cambridge).

Smolin, L. 1997, *The Life of the Cosmos* (Oxford University Press, Oxford).

Spengler, O. 1918, *Decline of the West* (1996 edition by Alfred A. Knopf Publisher, New York).

Steckline, V. S. 1983, *Am. J. Phys.* **51**, 894.

Tipler, F. J. 1981, *Q. J. R. astr. Soc.* **22**, 133.

Tipler, F. J. 1982, *Observatory* **102**, 36.

---

[1] Eddingtom (1928), p. 85. It is interesting to note that these words of Eddington preceded for more than two decades the coining of the expression "Big Bang", so they should not be interpreted as a critique of a particular model (after all, the first model which could, in a loose sense, be called a Big Bang model, was constructed by Lemaître only in 1931), but as rejection of the general concept of originating of the world in a finite moment of time.

[2] The most complete review of modern cosmological paradigm may be found in the monograph of Peebles (1993).

[3] For a sketch from the pen of a "contact pessimist" see Tipler (1981).

[4] It seems clear that Empedocles held a sort of the absolutist (substantivalist) theory of the nature of time. In particular, the fragment B 16 of the Diels collection reads (according to the translation of Burnet 1908): "For of a

truth they (Strife and Love) were aforetime and shall be; nor ever, methinks, will boundless time be emptied of that pair."

[5] Another pioneering contribution of Empedocles lies exactly in separation (the earliest one in Western thought!) of physical nature and artifacts of human cognizance. See, for instance, the Diels' fragment B8, reading (in Burnet's translation): "There is no coming into being of aught that perishes, nor any end for it in baneful death; but only mingling and change of what has been mingled. Coming into being is but a name given to these by men." Even more telling along the same lines are fragments B11 and B15.

[6] Fragment B 21 in Diels (1983), translation by Burnet (1908). Similar content can be found in B 20.

[7] This is clear, for instance, from the fragment A 10 in Diels (1983), preserved by Plutarch, in which it is explicitly asserted that formation and destruction of many worlds occurs within the global temporal infinity. In the continuation of the very same excerpt from *Stromateis*, an evolutionary doctrine is attributed to Anaximandros: "...Farther he says that at the beginning man was generated from all sorts of animals, since all the rest can quickly get food for themselves, but man alone requires careful feeding for a long time; such a being at the beginning could not have preserved his existence." (Fairbanks 1898) Hyppolites quotes Anaximandros as emphasizing the nature of *apeiron* as eternal (B 2), obviously in opposition to mankind, which has a fixed beginning in time. Even more intriguing is the doctrine ascribed to Anaximandros by Cicero: "It was the opinion of Anaximandros that gods have a beginning, at long intervals rising and setting, and that they are the innumerable worlds. But who of us can think of god except as immortal?" Did he have in mind essentially what we today denote as supercivilizations?

[8] In translation of William E. Leonard, available via WWW Project Gutenberg (Lucretius 1997).

[9] Of course this statement is not to be understood in the trivialized sense which is sometime ascribed to the anthropic thinking as a whole. It is possible not only to imagine a counterfactual **present** universe with no life in it (that is exactly the position usually taken in physical science where this counterfactual universe is identified with the real one, and any discrepancy is discarded as anthropocentrism, a useful if not always recommendable practice), but to imagine, and even calculate the properties of a future universe with no effects of life and intelligence. This anti-Dysonian approach is still present in physical eschatology and it may as well lead to useful and informative **approximations** (e.g. Adams and Laughlin 1997).